\documentclass[aps,pre,groupedaddress,notitlepage]{revtex4-1}[12pt]
\usepackage{graphicx,amsmath,amssymb}
\usepackage[usenames,dvipsnames]{xcolor}

\begin{document}

\title{Creep and flow of glasses: strain response linked to the spatial distribution of dynamical heterogeneities}


\author{T. Sentjabrskaja$^1$}
\author{P. Chaudhuri$^2$}
\author{M. Hermes$^3$}
\author{W. C. K. Poon$^3$}
\author{J. Horbach$^2$}
\author{S. U. Egelhaaf$^1$}
\author{M. Laurati$^{1}$}
\email[]{marco.laurati@uni-duesseldorf.de}
\affiliation{$^1$ Condensed Matter Physics Laboratory, Heinrich Heine 
University, Universit{\"a}tsstr. 1, 40225 D{\"u}sseldorf, Germany.\\ 
$^2$ Theoretical Physics II, Heinrich Heine University, Universit{\"a}tsstr. 1, 
40255 D{\"u}sseldorf, Germany.\\
$^3$ SUPA, School of Physics \& Astronomy, The University of Edinburgh, 
Mayfield Road, Edinburgh EH9 3JZ, United Kingdom.}

\date{\today}

\begin{abstract}
Mechanical properties are of central importance
to materials sciences, in particular if they depend on external stimuli.  Here we investigate the rheological
response of amorphous solids, namely colloidal glasses, to external
forces. Using confocal microscopy and computer simulations, we establish a quantitative link between the macroscopic
creep response and the microscopic single-particle dynamics.
We observe dynamical heterogeneities, namely regions
of enhanced mobility, which remain localized in the creep regime, but grow
for applied stresses leading to steady flow.
These different behaviors are also reflected in the {\it average}
particle dynamics, quantified by the mean squared displacement of the individual particles, and
the fraction of active regions. Both microscopic quantities are found to be proportional to the macroscopic
strain, despite the non-equilibrium and non-linear conditions during
creep and the transient regime prior to steady flow.
\end{abstract}

\maketitle


The properties of materials not only depend on their
chemical composition, but also on the arrangement and dynamics of
their constituents.  It is thus crucial to understand the link between
the macroscopic behaviour and the microscopic single-particle level.
The relation between an applied mechanical force and microscopic
processes is understood for crystalline, i.e.~ordered, materials.  Crystalline solids (like metals, ceramics
or minerals) irreversibly deform when subjected to a load which is
small enough to avoid fracture, although this response is very slow.
This kind of response is called creep and originates from the presence
of defects in the otherwise ordered arrangement of atoms. The diffusion
of vacancies and dislocations is responsible for the observed plastic
deformation \cite{poirier_book}. The same relation and microscopic processes cannot occur in amorphous,
i.e.~disordered, materials.

Nevertheless, in amorphous solid-like materials, a similar {\it macroscopic} creep response is
observed  under application of shear stresses below the yield stress
$\sigma_{\mathrm{y}}$, i.e.~below the transition from an elastic
to a plastic response.  The macroscopic creep response has been
intensively studied in metallic, polymeric and colloidal glasses
\cite{inoue_book,oswald_book,larson_book,laurati_jor,divoux11,pham08}.
Several models
\cite{mckenna94,sgr,falkstz1998,derec,siebenbuerger12,spaepen,argon,johnson}
successfully describe the time evolution of the strain measured
during creep, namely its characteristic sub-linear time dependence. However,
the relation of the creep response to the microscopic structure
and dynamics has hardly been determined and is not well understood.
Due to the disordered structure of amorphous solids the concept of defects
is not applicable and a microscopic mechanism different from the one in
crystalline solids must be responsible for creep. 
Thus, to make progress, microscopic observations on a single-particle level during creep tests are required.

Combining experiments and simulations, we investigated colloidal glasses
when constant stress is suddenly applied, i.e.~during creep tests. We reveal a quantitative link
between the macroscopic rheological response and the microscopic dynamics.
This is possible due to recent developments in simultaneously performing
rheology and confocal microscopy \cite{besseling09,blair}. During creep
flow near the yielding threshold, we observe that very few particles
undergo large non-affine displacements which leads to 
pronounced, but spatially localized, dynamical heterogeneities and 
sub-diffusive
dynamics. In contrast, for stresses beyond the yield stress, transient
super-diffusive dynamics mark the onset of steady flow. At the same time,
growing domains of enhanced dynamic activity are present, with their
number correlating with the macroscopic strain.  This is reflected
in a correlation between the macroscopic strain and the single
particle displacements. In addition to the steady-state flow regime,
this correlation also holds in the creep and transient states, specially
for stresses near the yield stress.  Hence, we can quantitatively relate
the macroscopic rheological response of soft glasses to the average and
heterogeneous microscopic dynamics which are spatially localized 
during creep but span the entire system at large stresses that lead to flow. 
The different microscopic behavior thus reflects the different macroscopic response during creep and flow, respectively.
This extends previous observations
to non-linear and non-equilibrium situations. Furthermore, 
as we observe the same behavior for different systems, realized 
in the experiments and simulations, this appears to be a general feature of glasses.

%
\section*{Results and Discussion}
In our experiments and simulations we investigated two different model colloidal
glasses. In the experiments, the glass is a binary mixture of
sterically stabilised PMMA spheres with a size ratio of 5, dispersed
in a density and refractive index matching solvent, with total volume
fraction $\phi=0.61$ and a relative volume fraction of small spheres
$x_{\mathrm{S}} = \phi_{\mathrm{S}}/\phi = 0.1$.  In this binary glass, 
the motion of the large particles is arrested via caging by neighbouring large particles  \cite{aip_proc,sentjabrskaja13,sentjabrskaja14}. In our molecular
dynamics simulations, the glass is formed by a binary
Yukawa fluid of large and small spheres with size ratio $1.2$ and
a relative number fraction of small spheres of 0.5, large enough to
prevent crystallization. This system is quenched to $T = 0.10$, i.e well below the mode-coupling
critical temperature of the system, $T_{\mathrm{c}} = 0.14$.  All times
are normalized; in the experiments by the short-time one-dimensional Brownian diffusion
time of the large spheres, $\tau_0^\mathrm{exp}
= 3\pi\eta d_{\mathrm{L}}^3/8k_{\mathrm{B}}T \approx 3.76$~s, where $d_{\mathrm{L}}$ is the diameter of the large spheres, $\eta$ is the viscosity and $k_{\mathrm{B}}T$ the thermal
energy, and in the simulations by the time unit $\tau_0^\mathrm{sim}=(m
d_{\mathrm{S}}^2/\epsilon_{\mathrm{SS}})^{1/2}$, where $m, d_{\mathrm{S}}$
and $\epsilon_{\mathrm{SS}}$ are units of mass, length and energy, respectively, with
$d_{\mathrm{S}}$ the diameter of the small spheres and $\epsilon_{\mathrm{SS}}$ the energy-scale corresponding to the interaction
between small particles. 
The colloidal glasses investigated in experiments and simulations hence involve different interactions and different mixing and size ratios  of their components. Using these different model systems allows us to explore the general features in the response of glasses to externally applied stresses.


\bigskip

\textbf{Macroscopic Strain is Quantitatively Related to Single-Particle
Mean Squared Displacement.} We performed a step to an applied constant
stress ($\sigma=$~const) on an initially quiescent glass. In the
experiments, the stress was applied  using a commercial stress-controlled
rheometer, while in the simulations one wall was pulled by a constant
force $F_0$.  We monitored the macroscopic response via the time
evolution of strain $\gamma(t)$. This situation is in contrast to
the case of imposing a constant shear-rate ($\dot\gamma=$~const)
\cite{varnik06,besseling07,schall07,zausch08,nick12,laurati12,brader09},
where the bulk stress $\sigma(t)$ is monitored. Unlike for an applied
shear rate $\dot\gamma$, when constant stress $\sigma$ is applied there is no
timescale imposed and flow regimes below yielding can be investigated.
The choice of control parameter, i.e.~constant $\sigma$ or constant
$\dot{\gamma}$, hence determines the intermediate flow states via which a
glass evolves from the quiescent state to steady flow \cite{derec}. In
the following, we exploit these possibilities and link the increasing
macroscopic strain to the evolution of local particle motions, using
stress as the external variable and including stresses below the yield
stress $\sigma_{\mathrm{y}}$.  In experiments, we estimated the yield
stress of the glass, $\sigma_{\mathrm{y}}\approx 0.010$~Pa, from the
stress at the crossing point of the storage and loss moduli in large
amplitude oscillatory shear measurements at 1~rad/s. In simulations,
at $T=0.10$ the yield stress $\sigma_{\mathrm{y}} = 0.072$ (in simulation units) was estimated by
strain-rate controlled simulations \cite{pinaki_condmat}.

\begin{figure}[tb]
\includegraphics[width=6in]{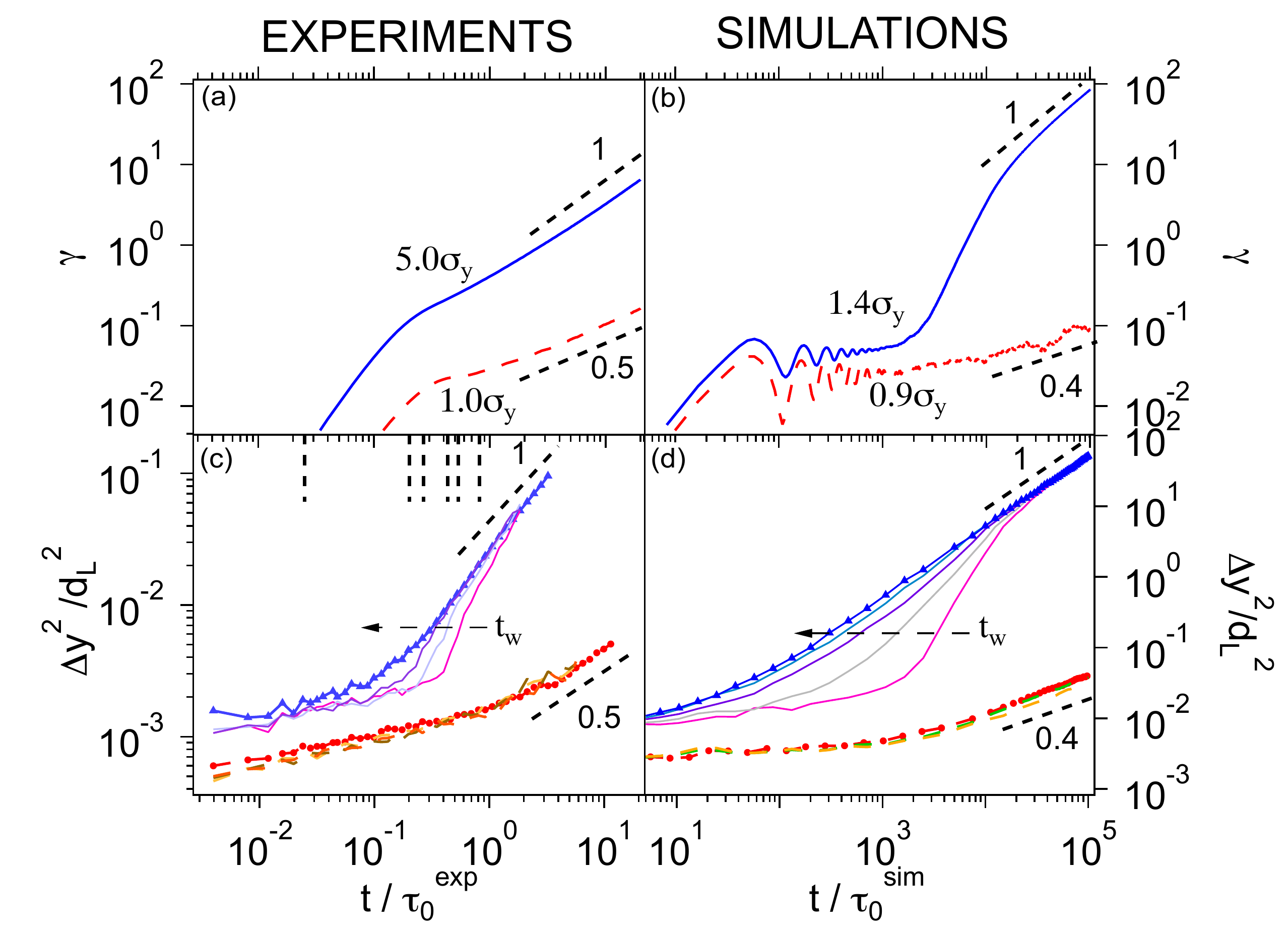}
\caption{\label{fig1}
Comparison of (left) experimental and (right) simulation results.
(top) Time-dependence of the strain $\gamma(t)$ for applied stresses
$\sigma$ as indicated, relative to the yield stress $\sigma_{\mathrm y}$.
(bottom) Mean squared displacement in the vorticity direction, $\Delta y^2$,
(indicated by same colors and line styles), immediately after stress
application, i.e.~for waiting time $t_{\mathrm{w}}=0$,  and larger
$t_{\mathrm{w}}$ (as indicated) until the steady-state is reached,
i.e.~$t_{\mathrm{w}} \to \infty$ (symbols). For the smaller applied stress, $\Delta y^2$ is divided by a factor 3 for clarity, both in experiments and simulations.}
\end{figure}

If the applied stress $\sigma \approx \sigma_{\mathrm{y}}$, a
characteristic creep response is observed with the strain increasing
sub-linearly with time within the experimental window, $\gamma \sim t^a$
with $a \approx 0.5$ (Fig.~\ref{fig1}a, broken line).  Furthermore, for
$\sigma = 0.9 \, \sigma_{\mathrm{y}}$ (Fig.~\ref{fig1}b, broken line), a
smaller effective exponent is found, in agreement with previous results
\cite{andrade1910,siebenbuerger12,pinaki_condmat,rosti10,divoux11}.
Hence, for $\sigma \lesssim \sigma_{\mathrm{y}}$, the deformation
occurs extremely slowly and the system is unable to reach a steady
state within the observation time.  This is reflected in the particle
dynamics in vorticity (neutral) direction, namely the mean squared
displacement (MSD) $\Delta y^2(t)$ (Fig.~\ref{fig1}c,d, broken lines).
In experiments and simulations, at short times the increase of the
MSDs is limited, consistent with caging, while at longer times a sub-diffusive regime is observed;
$\Delta y^2 \sim t^b$ with $b<1$. We find $b \approx a$ within the
explored time window.  The MSDs show little change with the waiting
times $t_{\mathrm{w}}$ after the beginning of the stress application
(Fig.~\ref{fig1}c,d, broken lines).  The observed macroscopic creep
response and the absence of steady-state flow is thus connected to the
particles' inability to diffuse.

In contrast, if $\sigma \gg \sigma_{\mathrm{y}}$, the strain response
shows a rapid transition to a steady flow regime, which corresponds to
$\gamma \sim t$, i.e.~$\dot{\gamma}$ is constant (Fig.~\ref{fig1}a,b,
solid lines).  The MSDs again display caging at intermediate times
(Fig.~\ref{fig1}c,d, solid line), while at long times diffusion
occurs \cite{nick12,koumakis:2013,mohan:2013,lin:2013}. The slightly lower MSD plateau observed in experiments is due to cage 
constriction \cite {aip_proc} and is also observed in Brownian dynamics
  simulations \cite {nick12,laurati12}, but not in molecular dynamics
  simulations where the microscopic dynamics is Newtonian. In between caging and long time diffusion, a
transient super-diffusive regime is observed. This coincides with the
transition of the rheological response from the initial elastic regime
to the flow regime. Note that in the experiments, the initial superlinear
  increase in strain is a known effect of the rheometer's inertia \cite
  {weitz08}. With increasing waiting time $t_{\mathrm{w}}$
(Fig.~\ref{fig1}c,d), super-diffusion occurs at increasingly earlier times
and for increasingly shorter time intervals, until it almost disappears in
the steady state.  Thus, the onset of flow is characterized by transient
super-diffusion and, subsequently in the steady state, by diffusion.
This indicates that the different regimes in the macroscopic strain
response $\gamma(t)$ are reflected in different features of the single-particle
dynamics, here characterized by the MSD $\Delta{y^2}(t)$.

We now quantitatively investigate the relation between the macroscopic strain $\gamma(t)$
and the microscopic MSDs $\Delta{y^2}(t)$.  In the case of steady flow
$\gamma(t) = \dot{\gamma}t$ and, since then the particles diffuse,
$\Delta{y}^2(t) \sim D(\sigma)t$, which implies that $\Delta{y}^2(t)
\sim [D(\sigma)/\dot{\gamma}]\gamma(t)=C(\sigma)\gamma(t)$. Previous
experiments and simulations under constant applied shear rate have found
$D\sim\dot{\gamma}^{0.8}$ \cite{varnik06,besseling07}, which implies
$C(\sigma)\sim \dot{\gamma}^{-0.2}$, since  stress and strain control
are equivalent in steady flow. In our case, in the asymptotic diffusive
regime (corresponding to $\gamma \gg 10$, Fig.~\ref{fig1}a,b) we
observe an approximate linear relation $\Delta{y}^2(t) \sim \gamma(t)$
for a large range of $\sigma$ (Fig.\ref{fig2}a, Sec.~1 in Supplementary Information).
The slight shifts between the curves for different $\sigma$ occur due to
the expected behaviour of $C(\sigma)$ (Fig.~\ref{fig2}b). If $\Delta{y}^2$
is rescaled by $C(\sigma)$, the data fall onto a single line of slope 1
(Fig.~\ref{fig2}c).

\begin{figure}[tbp]
\includegraphics[width=5in]{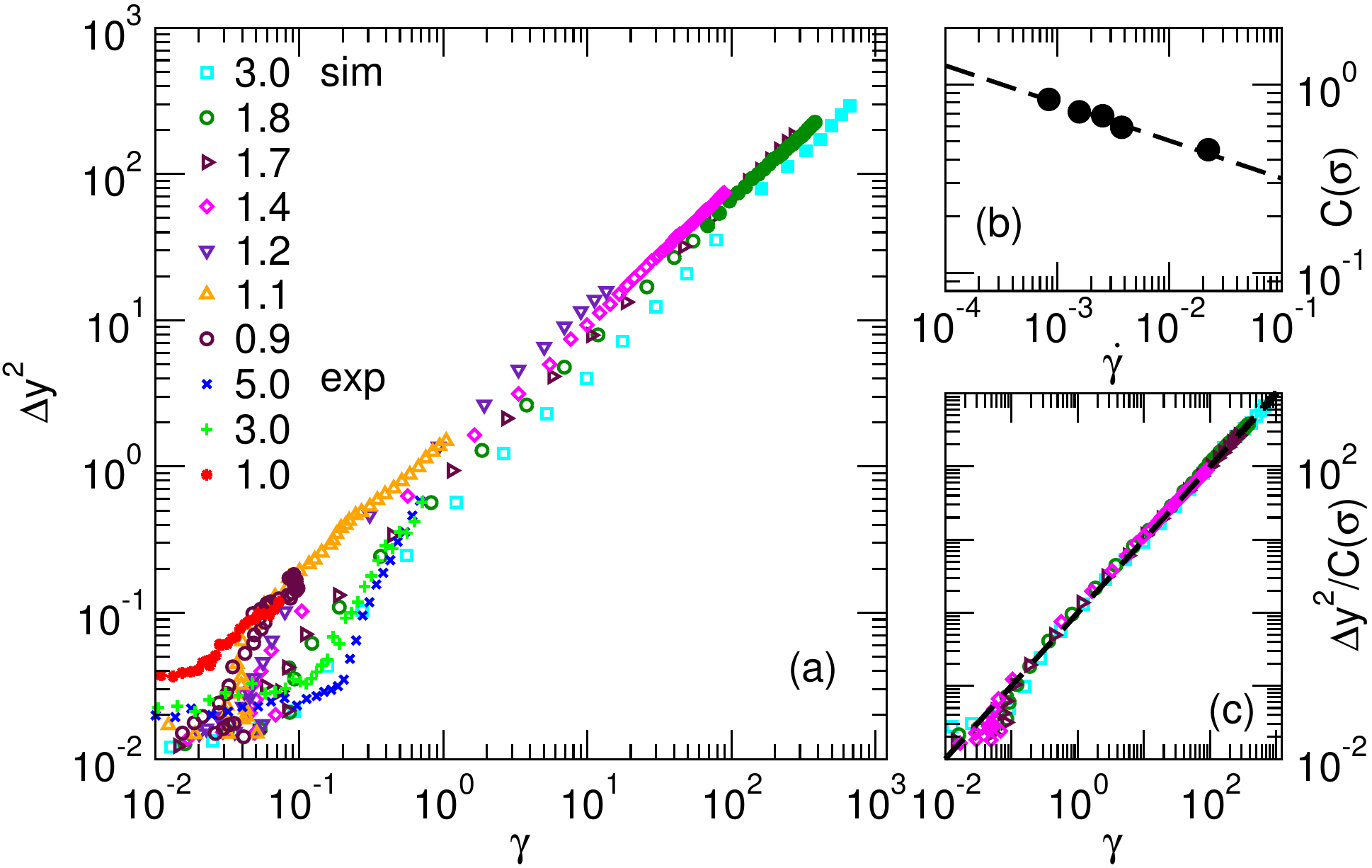}
\caption{\label{fig2} (a) Mean squared displacement in vorticity
direction, $\Delta{y}^2$, as a function of strain $\gamma$ for different
values of the applied stress $\sigma/\sigma_\mathrm{y}$ obtained in
experiments and simulations (Sec.~1 of Supplementary Information). The experimental
$\Delta{y}^2$ values are multiplied by a constant
factor in order to match the simulation data. (b) Ratio of diffusion
coefficient to shear rate, $C(\sigma)=D(\sigma)/\dot{\gamma}$, obtained
from fits (Sec.~1 of Supplementary Information), as a function of the shear-rate
in the steady state, $\dot{\gamma}$.  The dashed line indicates a
power-law $C(\sigma)\sim\dot{\gamma}^{-0.2}$. (c) Scaling plot of
$\Delta{y}^2/C(\sigma)$ as a function of $\gamma$, with the dashed
line indicating a slope of 1.}
\end{figure}

Although our argument for the relation $\Delta{y}^2(t) \sim
\gamma(t)$ is based on the assumption of steady flow, the relation surprisingly
also holds in non-steady states for $\gamma \lesssim 10$, which
corresponds to creep (for $\sigma \lesssim \sigma_{\mathrm{y}}$) or the
transient regime before steady flow (for $\sigma > \sigma_{\mathrm{y}}$).
In contrast, $\Delta{y}^2(t) \sim \gamma(t)$ does not hold for large
stresses $\sigma$ and small strains $\gamma$ (or short times $t$).
In both, experiments and simulations, systematic deviations are seen
to occur with increasing stress. The deviations occur due to a time
lag between the particles' motion beyond their cages and the onset of
macroscopic deformation (Fig.~SM-1, Supplementary Information).  Moreover,
at very short times, i.e.~in the initial elastic regime (Fig.~\ref{fig1}),
the proportionality is also not observed.  This suggests that the observed
correlation is a consequence of the plasticity that develops after the
initial elastic regime.  Our observations mark a clear difference between
the yielding response under applied constant stress (investigated here)
and applied constant shear-rate (investigated in \cite{zausch08,laurati12}). In the latter,
$\Delta{y}^2(t)\sim\gamma(t)$ cannot hold in the transient regime, where
$\Delta{y}^2(t)$ increases superlinearly with $t$ while $\gamma(t)$
increases linearly. Note that this connection between nonlinear
strain and the single-particle dynamics is an implicit assumption in a
recent theoretical approach based on a nonlinear 
Langevin equation \cite{schweizer2008,schweizer2010,schweizer2010r}. 
Our data indicates to what extent such a connection is valid.\\

\begin{figure}[tbp]
\includegraphics[width=4.8in]{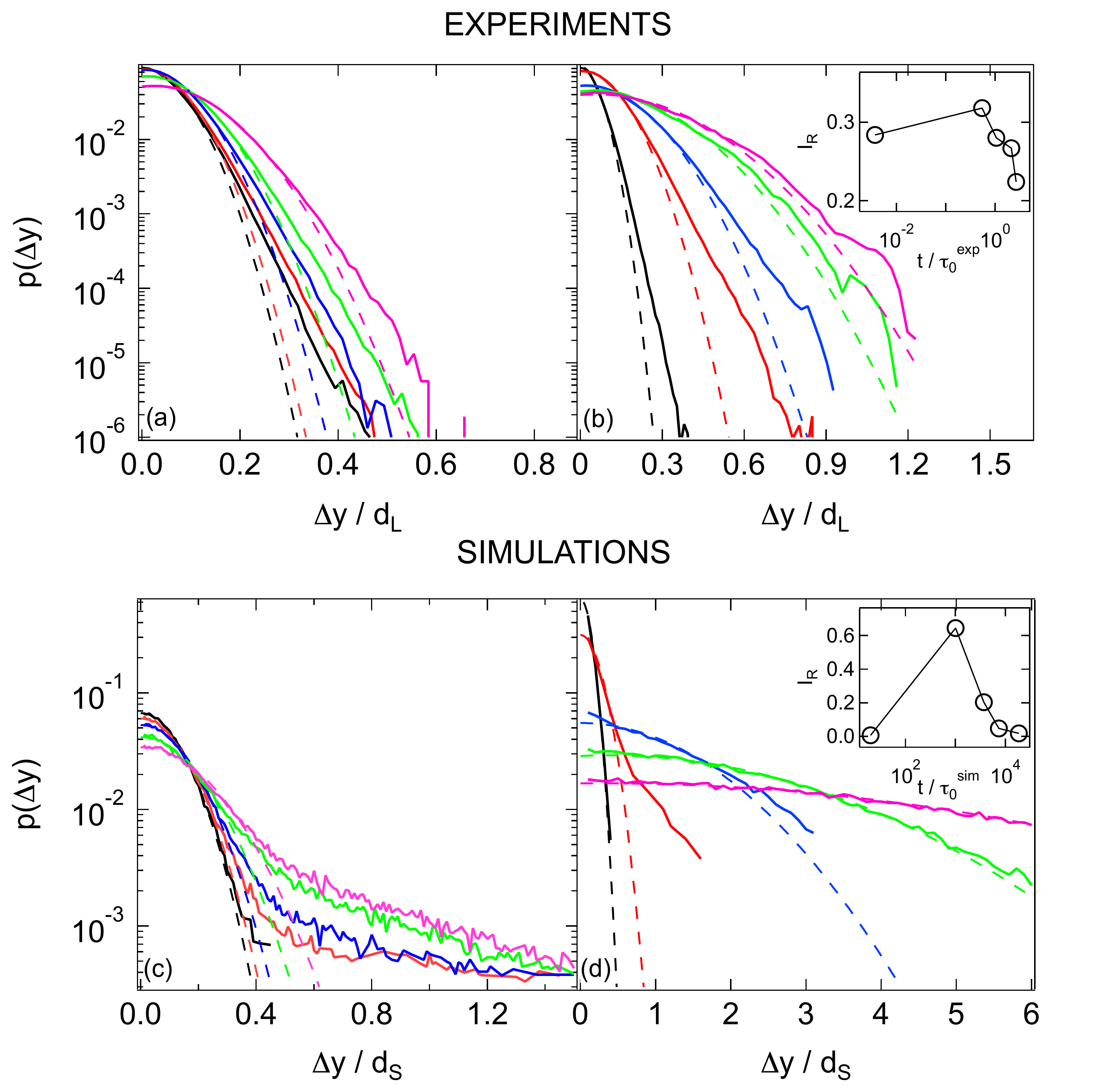}
\caption{\label{vanhove}Van Hove self-correlation functions,
i.e.~distributions of displacements $p(\Delta y)$, determined (top)
by experiments for (a) a stress $\sigma \approx \sigma_{\mathrm{y}}$,
a waiting time $t_{\mathrm{w}} = 0$ and times $t/\tau_0^\mathrm{exp}
= 0.0037$, 0.54, 1.05, 2.12 and 5.31 (left to right) and (b) $\sigma
\approx 5\sigma_{\mathrm{y}}$, $t_{\mathrm{w}}=0$ and same times,
except the longest time here is $t/\tau_0^\mathrm{exp} = 2.65$,
(bottom) by simulations for (c) $\sigma \approx 1.1\sigma_{\mathrm{y}}$,
$t_{\mathrm{w}} = 0$ and $t/\tau_0^\mathrm{sim} = 20, 3.7\times 10^3,
11\times 10^3, 56\times 10^3$ and $110\times 10^3$ (left to right)
and (d) $\sigma = 1.53\sigma_{\mathrm{y}}$, $t_{\mathrm{w}} = 0$ and
$t/\tau_0^\mathrm{sim} = 20, 1.01\times 10^3, 3.7\times 10^3, 7.4\times
10^3$ and $18.6\times 10^3$ (left to right).  Dashed lines represent
Gaussian fits to $p(\Delta y)$ for small $\Delta y$. Insets: normalized integral $I_\mathrm{R}$ of the residuals of the Gaussian fits in the main plots, as a function of time $t/\tau_0$.}
\end{figure}

\bigskip

\textbf{Displacement Distributions Indicate Small Fraction of Mobile
Particles.} In addition to the characterisation of the particle
displacements via the MSD, i.e.~a mean value, we have also investigated
the distribution of the displacements, namely the self part of the van
Hove function $p(\Delta y)$.  For $\sigma \approx \sigma_{\mathrm{y}}$
(Fig.~\ref{vanhove}, left), at all times the van Hove functions exhibit
a nearly Gaussian shape for small $\Delta{y}$, which corresponds to
localised particles, and moderate exponential tails which correspond to
large displacements of a small fraction of particles.  The non-Gaussian
tails only slightly change with increasing time. This indicates
that shear-induced delocalisation is a very slow
process. In particular, large displacements at the shortest time of the
measurement $t_0/\tau_0^\mathrm{exp}=0.0037$ hardly occur and therefore
macroscopic flow is delayed.

For $\sigma > \sigma_{\mathrm{y}}$
(Fig.~\ref{vanhove}, right) shear leads to a larger deviation from a
Gaussian distribution with a significant number of large displacements.
The deviation from Gaussian behavior 
was quantified by the time dependence of the integral $I_{\mathrm{R}}(t)$ of the residuals of the 
Gaussian fits to $p(\Delta y)$, which, for each fixed time, was normalised to the integral of the distribution (Fig.~\ref{vanhove}, inset). A non monotonic trend of $I_{\mathrm{R}}(t)$ is observed, with a maximum value during the intermediate superdiffusive regime.
At later times $I_{\mathrm{R}}(t)$ continuously decreases and eventually vanishes when diffusion sets in and a Gaussian distribution of displacements is
recovered.\\

\begin{figure}[tbp]
\includegraphics[width=3.2in]{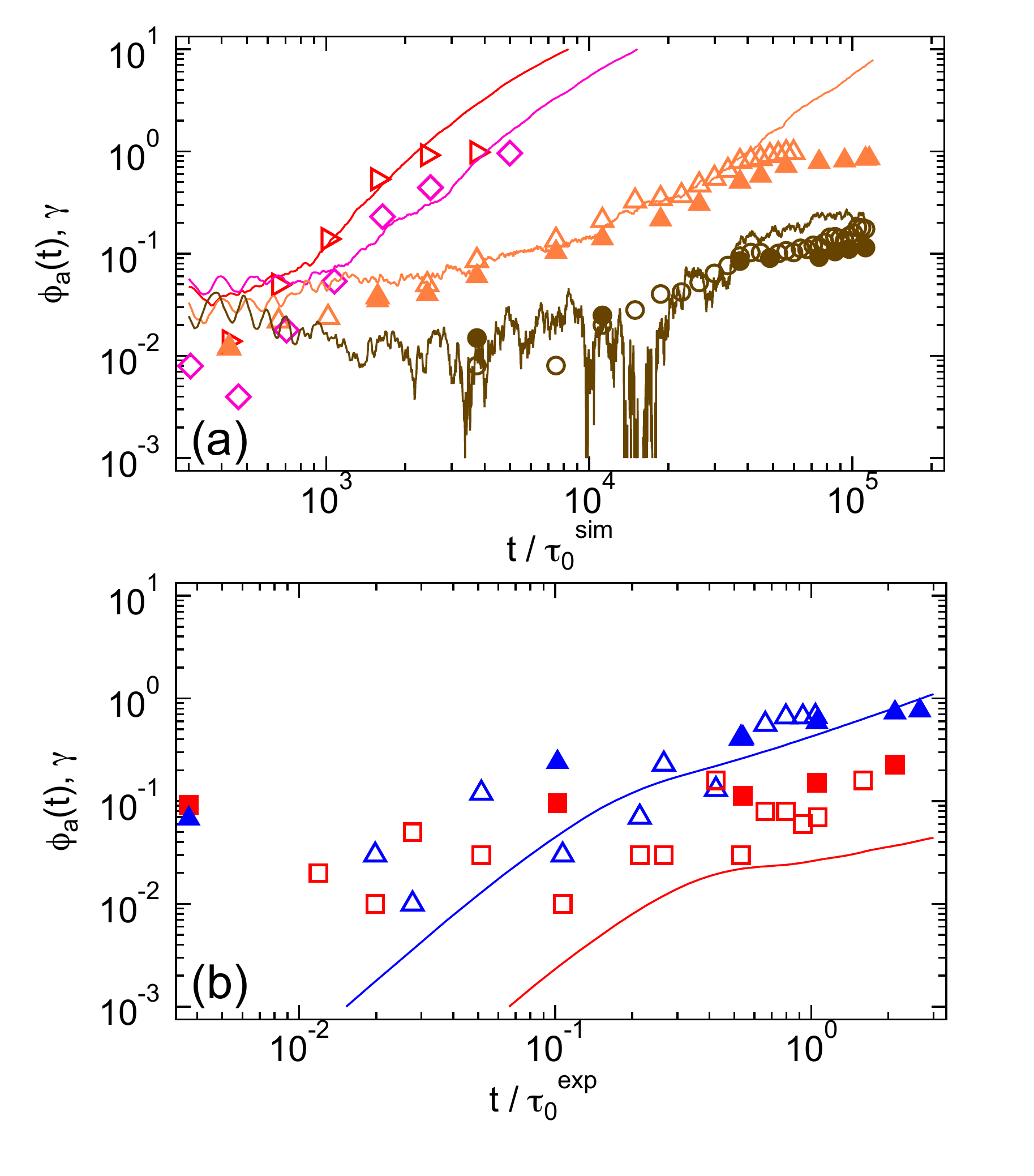}
\caption{\label{fig4} Fraction of active particles $\Phi^\mathrm{vH}_{\mathrm{a}}(t)$ (solid symbols) and active boxes $\Phi^\mathrm{b}_{\mathrm{a}}(t)$ (open symbols)
determined (a) by simulations at stresses $\sigma/\sigma_{\mathrm{y}}
= 0.9$ ($\bullet$, $\circ$), 1.1 ($\blacktriangle$, $\triangle$), 1.4 ($\lozenge$), 1.53 ($\triangleright$)
 and (b) by
experiments at $\sigma/\sigma_{\mathrm{y}}\approx 1.0$ ($\blacksquare$,$\Box$) and $5.0$
($\blacktriangle$,$\triangle$). Lines of the same colour represent the strain $\gamma$ for the corresponding
applied stresses, where the instantaneous strain is shown in the case
of the simulations. }
\end{figure}

\bigskip

\textbf{Evolution of Dynamical Activity Follows Macroscopic
Strain.} The
tails in the van Hove function $p(\Delta y)$ reveal the existence of a
small number of very mobile particles during the transient regime. 
We quantify the time evolution of the fraction of these mobile particles by the ratio $\Phi^\mathrm{vH}_{\mathrm{a}}(t)=I_{\mathrm{a}}(t)$
 with $I_{\mathrm{a}}$ the integral of $p(\Delta y)$ 
for displacements $\Delta y/d > 5\Delta y_{\mathrm{min}}$, with $d=d_\mathrm{S}$ and $d_{\mathrm{L}}$ in simulations and experiments, respectively (Fig.~\ref{fig4}, solid symbols). The value $\Delta y_{\mathrm{min}} = \sqrt{\Delta y^2(t_0)/d^2}$ is the localization length estimated from the MSDs at the shortest time
$t_0$ (Fig.~\ref{fig1}c,d). In simulations the time-dependence
of $\Phi^\mathrm{vH}_{\mathrm{a}}(t)$ closely follows that of the instantaneous
strain $\gamma(t)$, up to $\Phi^\mathrm{vH}_{\mathrm{a}}(t)=1$ (Fig.~\ref{fig4}a,
lines). In the experiments, similar results for $\Phi^\mathrm{vH}_{\mathrm{a}}(t)$
are observed (Fig.~\ref{fig4}b) except that, in contrast to the
simulations, $\gamma(t)$ is not the instantaneous strain but a time
average, leading to a small deviation between $\Phi^\mathrm{vH}_{\mathrm{a}}(t)$
and $\gamma(t)$. The macroscopic strain is therefore not only proportional to the average mean squared displacement (Fig.~\ref{fig2}) but also the fraction of mobile particles: this indicates that the mobile, dynamically active particles contribute most significantly to the mean squared displacement. This is true both below and above the yield stress.\\

\bigskip

\textbf{Spatial Distribution of Dynamical Activity is Heterogeneous.} We introduce spatial coarse-graining in order to reduce noise.
We divide the field of view into $10\times{10}$
square boxes, each with size $(2.8$~$d_{\mathrm{L}})^2$. For each
particle $i$, the displacement in the vorticity direction, $\Delta y_i
(t)= y_i(t)-y_i(t_0)$, was determined. The average particle mobility
in box $lm$, with $l$,$m = 1\dots 10$, was calculated according to
\begin{equation}
\mu_{lm}(t) = \left\langle \Delta y_i(t)\right\rangle_{lm}
\label{eq1}
\end{equation}
where $\left\langle \dots \right\rangle_{lm}$ denotes an average over all
the particles which were in the box $lm$ at $t = t_0$.  
A box $lm$ is defined active at time $t$ if $\mu_{lm}
> 5\Delta y_{\mathrm{min}}$, following the same criterion used to distinguish largest displacements of single particles in the van Hove functions (Fig.~\ref{fig1}c,d).
The fraction
of active boxes, $\Phi^\mathrm{b}_{\mathrm{a}}(t)=N_{\mathrm{a}}(t)/N_\mathrm{tot}$,
with $N_\mathrm{a}$ the number of active boxes and
$N_{\mathrm{tot}}$ the total number of boxes.
With time $\Phi^\mathrm{b}_{\mathrm{a}}(t)$ grows as the fraction of the single mobile particles $\Phi^\mathrm{vH}_{\mathrm{a}}(t)$  (Fig.~\ref{fig4}, symbols). Thus, the time-dependence of $\Phi^\mathrm{b}_{\mathrm{a}}(t)$ is also proportional to $\gamma(t)$. 
 A similar connection between the number of  active
regions and strain growth was experimentally observed in the creep
flow of frictional granular particles \cite{amon2012}. 
%
\begin{figure}[tbp]
\includegraphics[width=4.5in]{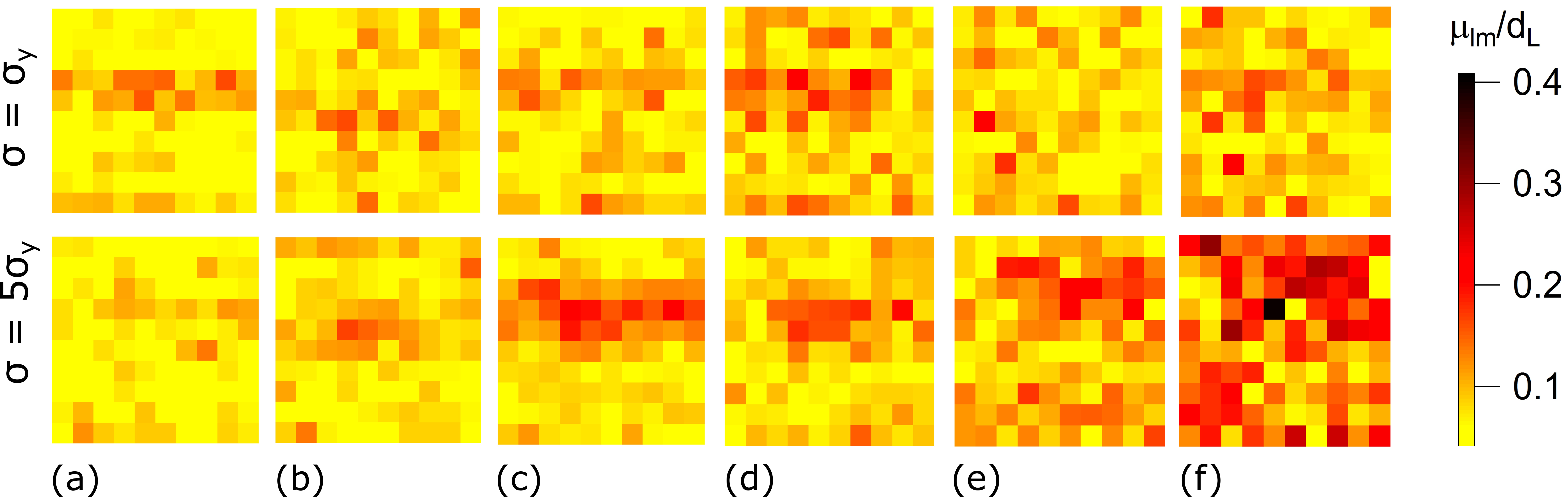}
\caption{\label{fig3} Maps of average particle mobilities $\mu_{lm}(t)$
within boxes $lm$ (Eq.\ref{eq1}) for (top) stress $\sigma\approx
\sigma_{\mathrm{y}}$ and (bottom) $\sigma\approx 5\sigma_{\mathrm{y}}$
and times $t/\tau_0^\mathrm{exp}=0.025, 0.20, 0.27, 0.43, 0.53, 0.80$
(a--f, indicated in Fig.~\ref{fig1}c by dashed lines) as observed in
experiments. The box size is $(2.8~d_{\mathrm{L}})^2$.}
\end{figure}

To investigate the existence of heterogeneity in the dynamical activity, we consider the spatial distribution of active boxes.
For $\sigma\approx\sigma_{\mathrm{y}}$, the distribution of local mobilities within
the velocity--vorticity plane does not indicate any prominent features (Fig.~\ref{fig3},
top). At any specific time, there are some active boxes with larger
mobilities, but the locations of the boxes with the largest mobilities
vary randomly with time. For $\sigma\approx 5\sigma_{\mathrm{y}}$,
similar mobilities occur at short times, when the localisation plateau
in the MSD is observed (Fig.~\ref{fig3}a,b, bottom).  In contrast,
at $t \gtrsim 0.3\tau_0^\mathrm{exp}$, roughly coincident with the
onset of super-diffusion in the MSDs determined for $t_{\rm w} =
0$ (Fig.~\ref{fig1}c), a region with enhanced mobilities emerges
(Fig.~\ref{fig3}c,d, bottom), expands with time (Fig.~\ref{fig3}e,
bottom) and spans almost the whole field of view once the system flows
(Fig.~\ref{fig3}f, bottom). Hence, the onset of flow (Fig.~\ref{fig1}a,b)
coincides with the appearance of a region of higher local mobility (Fig.~\ref{fig3}) and super-diffusive dynamics (Fig.~\ref{fig1}c,d). 
Furthermore, it leads to the pronounced non-Gaussian tails in the van Hove correlation function
at intermediate times (Fig.~\ref{vanhove}), which disappear once steady flow has developed and the dynamics again becomes more homogeneous (Fig.~\ref{vanhove}, inset).

The enhanced local mobilities do not result from sudden large
displacements, but occur through the accumulation of only
slightly above-average local, non-affine particle displacements. This has
been confirmed by calculating the instantaneous mobilities
from $0.18\tau_0^\mathrm{exp}$ to $0.46\tau_0^\mathrm{exp}$ and
$0.28\tau_0^\mathrm{exp}$ to $0.56\tau_0^\mathrm{exp}$, i.e.~for $10$
sampling times, instead of starting from the shortest measurement
time (as in Fig.~\ref{fig3}). No large instantaneous mobilities and
no significant difference to $\sigma\approx\sigma_{\mathrm{y}}$ are
observed (Sec.~2 in Supplementary Information). Similar results are obtained
in our simulations.  The occurrence of correlated plastic events
\cite{bocquet_fluidity,chikkadi2011} and avalanche-like behavior
\cite{lemaitre2010,procaccia2010} have been proposed as mechanisms driving
the onset of flow.  Such cooperative events might be connected to the
correlated local mobilities and their spreading observed in our study. 
The observed intermittency in the displacements might also be related to stick-slip motion \cite{gee1990}.

\begin{figure}[tbp]
\includegraphics[width=3.2in]{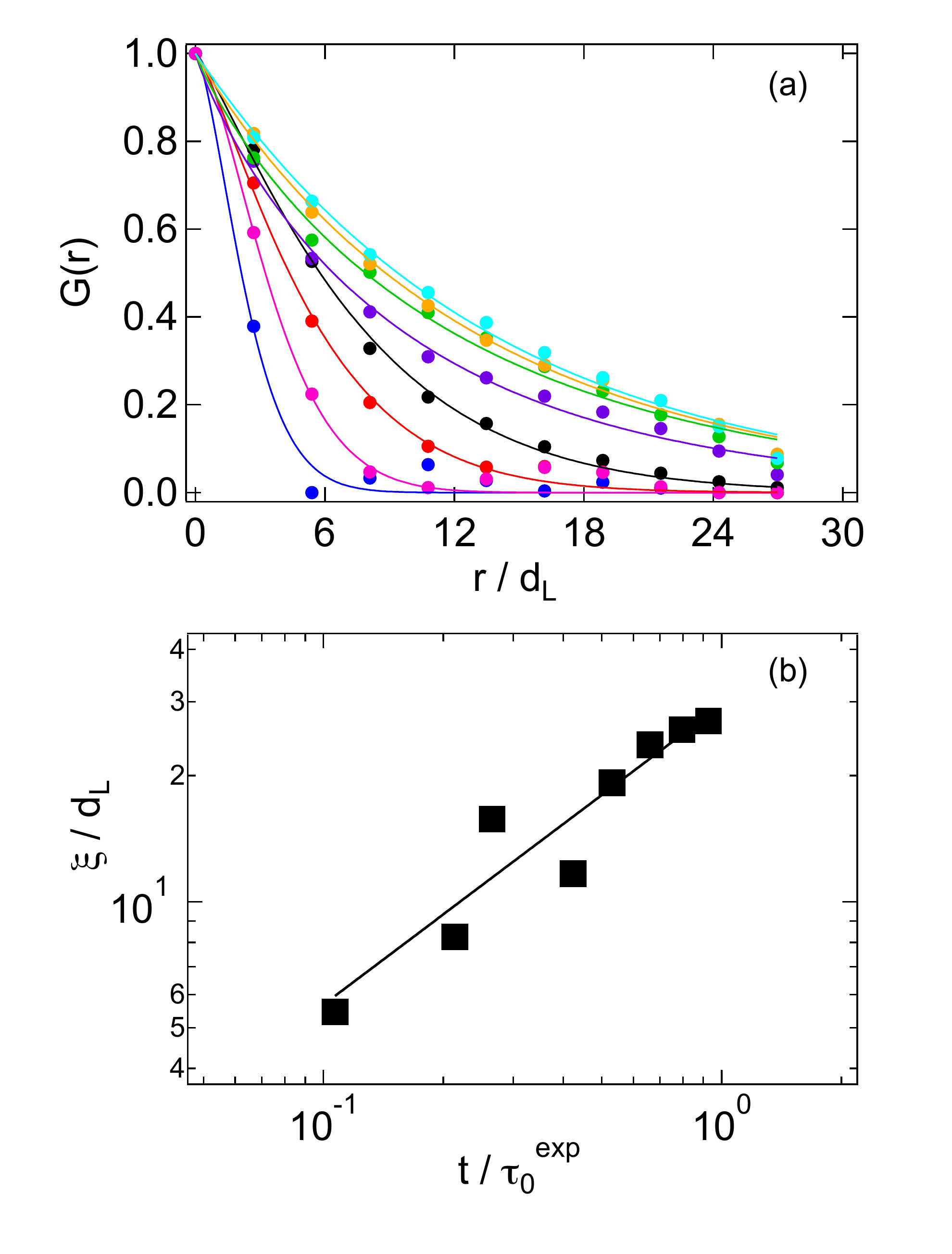}
\caption{\label{fig2_supp} (a) Box-box correlation functions
$G(r)$ for stress $\sigma/\sigma_{\mathrm{y}} \approx 5$  and time
$t/\tau_0^\mathrm{exp} = 0.1$, 0.20, 0.27, 0.43, 0.53, 0.66, 0.80 and 0.93
(left to right) as observed in experiments. Lines represent stretched
exponential fits. (b) Correlation length of active boxes, $\xi$, as a
function of time for $\sigma/\sigma_{\mathrm{y}}\approx 5.0$; the line
indicates $\xi/d_{\mathrm{L}}\sim t^{2/3}$.}
\end{figure}

The larger area in the velocity-vorticity plane monitored in the
experiments allows us to quantitatively investigate the spatial growth of
active regions.   
If the box $lm$ is active or inactive,
$n_{lm}$ is defined as $1$ or $0$, respectively.  Based on this
definition, we calculate the spatial correlation of active boxes,
that is the box-box correlation function, $G(r) = \langle n_{lm}
n_{l^{\prime}m^{\prime}}\rangle$ with $r^2 = (l{-}l^{\prime})^2 +
(m{-}m^{\prime})^2$ (Fig.~\ref{fig2_supp}a). The brackets $\langle \dots
\rangle$ indicate an average over the individual boxes. The characteristic
length $\xi$ of the spatial correlation $G(r)$ was determined by
fitting a stretched exponential function $f(r)=A\exp[-(r/\xi)^{\beta}]$
to $G(r)$. The correlation length $\xi(t)$ increases from an initial
value $\xi\approx 5d_{\mathrm{L}}$  at $t = 0.10\tau_0^\mathrm{exp}$
to $\xi\approx 30d_{\mathrm{L}}$ at $t = 0.92\tau_0^\mathrm{exp}$,
with $\xi(t)\sim t^{2/3}$ (Fig.~\ref{fig2_supp}b). 
For $\sigma \approx \sigma_\mathrm{y}$ the correlation length $\xi(t)$ instead does not grow and stays approximately constant for all times $t$ (data not shown).\\

\section*{Conclusions}
Using experiments and simulations, we demonstrated that  under applied
stress, the macroscopic deformation of glasses can be linked in a
consistent way to the single particle displacements. In particular, the strain is approximately linearly 
related to the single-particle MSD even in the time-dependent non-linear 
response regime, including the creep and the transient regime preceding steady flow. 
Furthermore, the fraction of active particles in the van Hove function as well as  the fraction of
active regions, i.e.~of groups of particles, is also proportional to the macroscopic strain.
Heterogeneities in the location of these active particles are present both for applied stresses smaller 
and larger than the yield stress. The spatial distribution of regions with larger displacements determines the onset of flow. For applied stresses
around the yield stress, i.e.~during creep, localised regions of enhanced
dynamical activity allow only for sub-diffusive dynamics. Increasing the stress
beyond the yield stress, the active regions grow heterogeneously and super-diffusive
transients emerge leading to particle diffusion  with steady flow
setting in. 
We observe qualitatively the same behavior for the different models studied
     in our experiments and simulations and thus expect that our observations represent generic features of glasses.


Future work should focus on understanding how the external stress leads
to the occurrence of locally enhanced mobilities, e.g.~whether these are
related to thermally activated local structural changes. Furthermore,
the mechanisms that drive the spreading of the active regions within
the plane as well as in the transverse direction need to be identified,
thereby providing possible links to transient shear banding  in the
velocity-gradient plane \cite{pinaki_condmat,moorcroft2013}. All these
would help to develop a more complete scenario for the fluidisation of
glassy systems under applied stress. Furthermore, it can open the route to the rational design of materials with desired response to applied stresses.

\section*{Methods}

\textbf{Experiments.} We investigated a mixture of sterically
stabilized PMMA spheres of diameters $d_{\mathrm{L}} = 1.76$~$\mu$m
(fluorescently labeled) and $d_{\mathrm{S}} = 0.35$~$\mu$m,
dispersed in a cis-decalin/cycloheptyl-bromide  mixture which
closely matches their density and refractive index.  After addition
of salt(tetrabutylammoniumchloride), this system presents hard-sphere like
interactions \cite{yethiraj03,Poon/Weeks/Royall}. The total volume
fraction is $\phi=0.61$ and the relative fraction of small spheres
$x_{\mathrm{S}} = \phi_{\mathrm{S}}/\phi = 0.1$.  
The formation of a glassy state in this mixture was demonstrated 
by using rheology and confocal microscopy measurements of the dynamics of large particles \cite{aip_proc,sentjabrskaja13,sentjabrskaja14}.
The presence of small spheres, with their larger energy density, increases the yield stress of
the system, thereby improving the quality of the rheological data while
still allowing for the simultaneous observation of the large spheres with
confocal microscopy \cite{besseling09,blair}.  The rheological
and confocal microscopy measurements reported in the manuscript were
obtained using a combination of a commercial MCR-301 WSP stress-controlled rheometer
(Anton-Paar) and a VT-Eye confocal unit (Visitech) mounted on a Nikon Ti-U
inverted microscope, with a Nikon Plan Apo 60x oil immersion objective (NA = 1.40). We
used a cone-plate geometry of diameter 50~mm, cone angle 1$^{\circ}$
and truncation gap 100~$\mu$m. The bottom plate consists of a microscope
coverslip which was coated with a mixture of PMMA particles of radius
0.885~$\mu$m and 0.174~$\mu$m. The surface of the cone is sandblasted. The
roughness of the geometries prevents wall-slip, as verified by imaging.
A solvent trap was used to reduce solvent evaporation.  Due to the
fact that rheological measurements on colloidal glasses are affected by
loading effects, shear history and aging, before each test a renjuvenation
procedure was performed in order to obtain a reproducible initial state
of the system.  After loading, we performed a dynamic strain sweep to
estimate the yield strain $\gamma_\mathrm{y}$ of the system from the
crossing point of the strain-dependent storage, $G'$, and loss, $G''$,
moduli.  Oscillatory shear at $\gamma = 3 \gg \gamma_\mathrm{y}$
was applied to induce flow and maintained until the time-dependent $G'$
and $G''$ reached a stationary state, typically after 200~s.  Afterwards,
oscillatory shear in the linear viscoelastic regime, $\gamma = 0.001$,
was applied until $G'$ and $G''$ became stationary, typically for $t
> 300s$.  The state characterised by the stationary values of $G'$ and
$G''$ was the initial state, prepared before each creep measurement. The
relative error on the strain determination during creep measurements
is smaller than 1\%.

Confocal microscopy images were acquired in a
velocity-vorticity plane about 6~mm from the center of the geometries
and 30~$\mu$m from the bottom plate.  Images with 512$\times$512 pixels,
corresponding to 51~$\mu$m $\times$ 51~$\mu$m, were acquired at a rate
of 67 frames per second, which ensured accurate particle tracking 
even at the highest applied stresses (typical movies in Supplementary Information). 
By imaging the truncation gap of the cone, we verified that bending of the 
coverslip is negligible \cite{besseling09}. This is also indicated by the fact that, 
despite the applied stress, 
the particles in the imaging plane remain perfectly in focus (movies in Supplementary Information). 
The fact that we can image the 
truncation gap of the cone is also used to check that the bottom plate is perpendicular 
to the rotation axis of the cone. Particle coordinates and trajectories were
extracted from the images using standard routines \cite{crocker96}.
Mean squared displacements from four independent measurements were
averaged. The noise contribution to our MSD data was estimated from
the MSD of an immobile sample, resulting in $\Delta y^2/d_\mathrm{L}^2
\approx 4\times 10^{-4}$, i.e.~a factor of about 2.5 times smaller than
the $\Delta y^2(t)$ values measured at short times.

\textbf{Simulations.} In our molecular dynamics simulations,
a 50:50 binary Yukawa fluid of large and small spheres with
size ratio $1.2$ is
investigated. The model parameters have been reported earlier
\cite{zausch08,zausch09,pinaki_condmat}.  Our simulations have
been performed for samples consisting of $N=12800$ particles
and having dimensions $L_{\mathrm{x}}=26.66d_{\mathrm{S}},
L_{\mathrm{y}}=13.33d_{\mathrm{S}}, L_{\mathrm{z}}=53.32d_{\mathrm{S}}$.
We work in the $NVT$ ensemble using periodic boundary conditions,
the temperature being controlled by a Lowe thermostat \cite{lowe}.
The mode-coupling critical temperature of the system is $T_c = 0.14$. The
system is equilibrated at $T = 0.15$ and then instantaneously quenched
to $T = 0.10$, where it is aged for $10^4$~$\tau_0^\mathrm{sim}$.
At this time, the walls are generated by freezing particles at $0 <
z < 2d_{\mathrm{S}}$ and $L_{\mathrm{z}}{-}2d_{\mathrm{S}} < z <
L_{\mathrm{z}}$ \cite{pinaki_condmat}.  Stress is applied by pulling
one wall at a constant force $F_0$ in the $x$ direction.
For each applied stress, runs over $24$ independent replicas of the system
were averaged. Similar to the experiments,  the dynamics were measured
in a slice at the centre of the volume having thickness $13.3d_{S}$
and distance about $18d_{S}$ to the walls on each side.


%

\section*{Acknowledgements}
We thank A.B. Schofield for the colloidal particles and K.J. Mutch for help with the analysis of the experimental data. 
We acknowledge support from the Deutsche Forschungsgemeinschaft through the Research unit FOR1394 (Projects P2 and P8) 
and EU funding through the FP7-Infrastructures "ESMI" (CP$\&$CSA-2010-262348). 
Further, NIC J\"ulich is thanked for providing computing time. The Edinburgh work was funded by EPSRC grant EP/J007404/1.

\section*{Author contributions}
T.S., M.H., W.C.K.P., S.U.E and M.L. performed, analysed or interpreted the experiments which are based on equipment in the lab of M.H. and W.C.K.P.. P.C. and J.H. performed, analysed or interpreted the simulation. All authors contributed to the interpretation and comparison of the data as well as the writing of the manuscript.

\section*{Additional Information}
Competing financial interests:
The authors declare no competing financial interests.




\end{document}